\documentclass{ws-procs961x669}            
\begin{document}
\title{Inflationary electroweak/particle phase transitions\\
and new classical gravitational waves on CMB \footnote{This talk is mainly based on the reference~\cite{Jiang:2015qor}.}\\
}

\author{Sichun Sun}

\address{Institute for Advanced Study, Hong Kong University of Science and Technology,\\ Clear Water Bay, Hong Kong,\\
$^*$E-mail: sichun@ust.hk\\
http://www.ust.hk/}

\begin{abstract}
 Particle phase transitions in the early universe including electroweak and grand unification ones are well-studied subjects. We point out that there are new possible particle phase transitions around inflation. Those new inflationary particle phase transitions, if of the first order, may yield low-frequency gravitational waves (GWs) due to bubble dynamics, leaving imprints on the cosmic microwave background (CMB). In contrast to the nearly scale-invariant primordial GWs caused by vacuum fluctuation, these bubble-generated GWs are classical and have scale dependent B-mode spectra. If decoupled from inflaton, the electroweak phase transition during inflation may serve as a mirror image of the one after reheating where the baryon asymmetry could be generated via electroweak baryogenesis (EWBG). The second new electroweak phase transition may also be the source for EWBG. 
 
\end{abstract}


\bodymatter
\section{Introduction}

Particle phase transitions happened in the early universe are rich subjects closely tied to beyond the standard model physics and particlegenesis in the early universe. With the advent of LHC and the discovery of Higgs, the questions around the electroweak phase transition have intensified. The direct detection experiments for GWs and collider searches are two major ways to probe particle phase transitions.

 Inflation is the leading paradigm in the very early cosmology to provide the seeds for CMB fluctuations. In the old inflation scenarios of the early days, the inflaton went through a tunneling process similar to a strong first order phase transition, to drive inflation to happen. This inflaton phase transition scenario is however not successful because the phase transition bubbles cannot collide. If one goes beyond the single field inflation scenario (which is actually a natural step after considering the presence of the standard model Higgs and many popular beyond the standard model scalar extensions\cite{Hamada:2015skp}), the phase transitions of the other fields during inflation remain valid. The energy density during inflation is very high in almost all inflation models to drive inflation, whereas the temperature during inflation is roughly a constant, so-called the Gibbons-Hawking temperature  $T_{GH}=H/(2\pi)$, which can be as low as $10^{-24}$GeV. One can find the experimentally allowed inflation models with different Hubble constant $H$ in Fig.~1.

We would like to stress here that our scenario is very different from Higgs inflation \cite{Bezrukov:2007ep,Hamada:2015skp} and Coleman-de Luccia bubble phase transitions\cite{Coleman:1980aw}. The energy density difference inside and outside the phase transition bubbles is not too large in our scenarios, so as inflation can happen both inside and outside bubbles. This way bubbles can collide and stir up turbulence in the thermal plasma to generate gravitational waves. 

The temperature contribution from radiation drops quickly at the beginning of inflation when all the particle contents get diluted, left with only contribution from curvature, $T_{GH}$. 
  With the lower scale inflation models, $H \sim T_{GH} \leq 10^2$GeV, the Higgs field/phase transition field stay at the origin of the potentials and tunnel to the spontaneous breaking phase due to the decreasing of the temperature, See Fig.2. Notice that even when $H\leq 10^2 $GeV, the energy density in that early universe is still as high as $\rho \sim (10^{10} \text{GeV})^4$. Moreover, the GUT phase transition, if exists, will happen in almost all the inflation scenarios.

The normal phase transitions happened in particle physics after reheating produce gravitational waves that direct detection experiments are currently looking for by Advanced LIGO, Advanced Virgo and LISA. The particle phase transitions here happened \textit{around  inflation} in the much earlier universe. They left an imprint on much larger scales in the universe today. If we assume a strong first order phase transition in particle physics (for electroweak phase transition it is preferred by baryogenesis), the drastic bubble nucleation process produced in phase transition will produce gravitational waves. This type of gravitational waves, produced around the same time as the primordial gravitational wave, but having a different yet scale-dependent spectrum, characterized by the bubble sizes at the time of collisions.  

\begin{figure}[h]
\begin{center}
\includegraphics[width=2.5in]{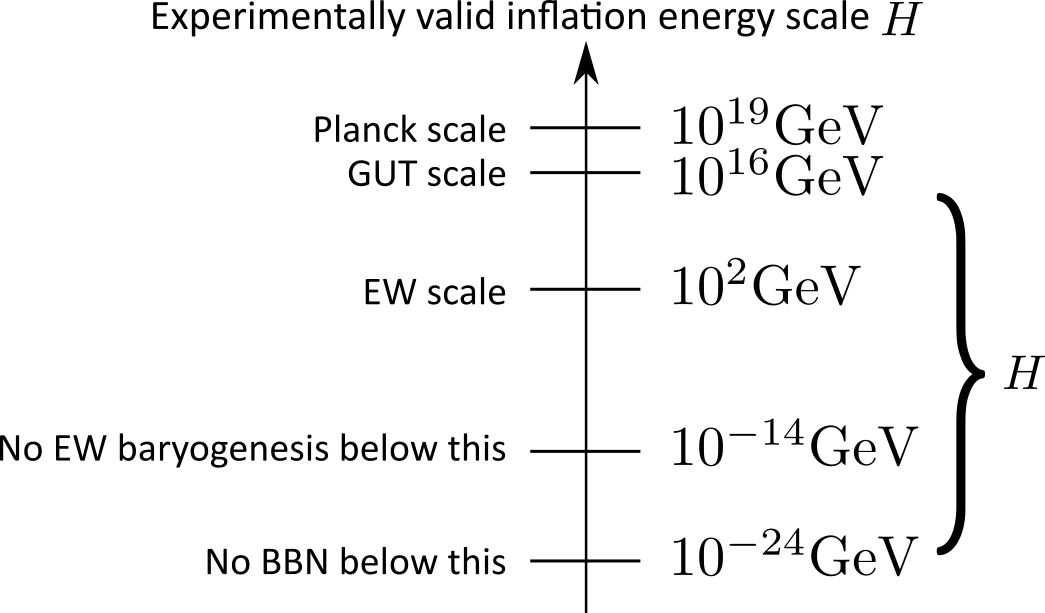}
\end{center}
\caption{In various inflation models, Hubble constant during inflation can take different values from $10^{-24}$ GeV up to $10^{14}$GeV. The upper bound is set by the latest experiment. When the Hubble constant is below $10^{-14}$ GeV, the universe can not reheat above $100$GeV later,  where the EW baryogenesis can be hardly achieved. Below $10^{-24}$ GeV, the reheated universe is too cool to have big bang nucleosynthesis.}
\label{aba:fig1}
\end{figure}

\begin{figure}[h]
\begin{center}
\includegraphics[width=2.5in]{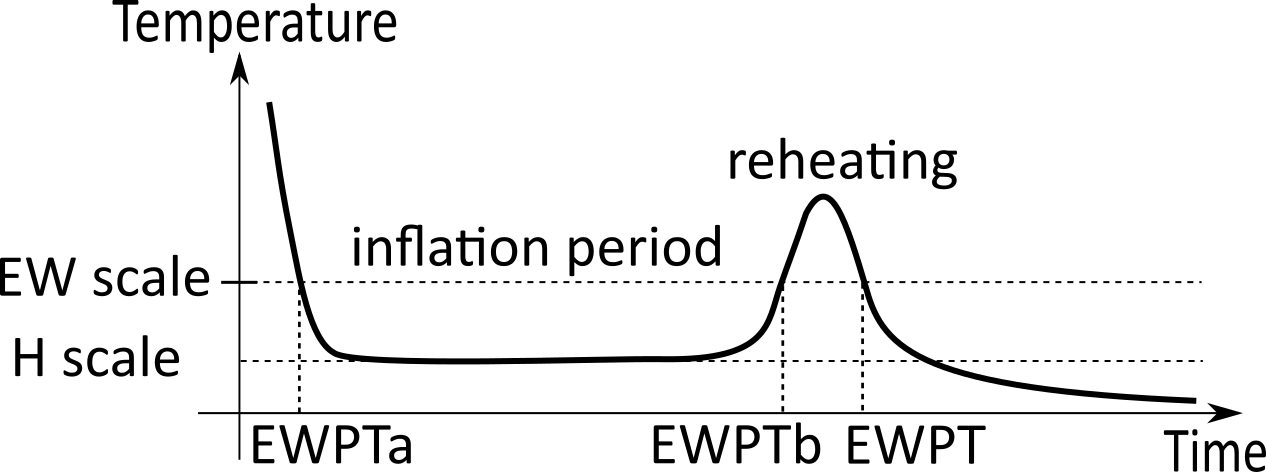}
\end{center}
\caption{The thermal temperature vs time in the early universe. Notice that the temperature during inflation comes mainly from the curvature contribution of the de sitter space, which is proportional to the Hubble constant. Two more electroweak phase transitions are proposed here, namely EWPTa and EWPTb.}
\label{aba:fig1}
\end{figure}

\section{Gravitational wave spectra}

We derive the power spectrum $P_\gamma$ coming from scale dependent GWs produced in de-Sitter space here. During inflation, the action can be written as two independent copies of probe field $\gamma_s$,
\begin{equation}
S=\sum_{s=+,\times}\int  d t  \frac{d^3 \bm k}{(2\pi)^3} a^3\Big[
 \frac{1}{2}\dot{\gamma_s }(\bm k )\dot{\gamma_{s }} (-\bm k ) - \frac{k^2}{2a^2} \gamma_s(\bm k ) \gamma_{s }(-\bm k) \Big]~.
\end{equation}

We can then quantize the $\gamma_s$ fields:
$
\gamma_s(\bm k,\tau)= v_k(\tau)a_{\bm k s}+v^*_k(\tau)a^\dagger_{-\bm k s }~.
$
Solving the equation of motion,  we can get the mode function $v_k$:
\begin{equation}
v_{k}(\tau)=\frac{H}{\sqrt{2k^3}}\Big(c_1( k) (1+ik\tau)e^{-ik\tau}+c_2( k) (1-ik\tau)e^{ik\tau} \Big) ~,
\end{equation}
where the coefficients $c_1(k),c_2(k)$ are subject to the consistency condition of quantization
$ |c_1|^2-|c_2|^2=1 $.

 The energy density of GWs is essentially the Hamiltonian density, given by
\begin{equation}
 \rho_{GW}=\int \frac{ dk}{k} \frac{ k^3}{2\pi ^2} \Big(  |\dot{v}_{k} |^2 +\frac{k^2}{a^2}|v_{k} |^2  \Big)~.
\end{equation}

Furthermore, the gravitational energy spectrum can be obtained as
\begin{eqnarray}
\Omega_{GW}(k,\tau)&=&k\frac{d\rho_{GW} }{dk}/ \rho_{\text{tot}} \\
&=&\frac{1 }{3H^2  M_p^2} \frac{ k^3}{2\pi ^2}
   \frac{|v_{k}\rq{}(\tau)|^2 +k^2|v_{k}(\tau)|^2 }{a(\tau)^2}~,
\end{eqnarray}
where we assume that the universe is spatially flat, meaning that $\rho_{\text{tot}}=\rho_{\text{critical}}=3H^2M_p^2$.  Particularly note that during inflation $\rho_{\text{tot}}=\rho_{\text{inflaton}}+\rho_{\text{rad}}+\rho_{\text{higgs}}$.

We can also calculate the power spectrum of GWs which measures the two point correlation: 
$
P_\gamma(k,\tau)=\frac{4k^3}{\pi^2 M_p^2} |v_k(\tau)|^2 ~.
$
This power spectrum of GWs contributes both temperature fluctuations and polarizations on the CMB.

We consider the classical limit for these GWs, where $c_1\approx  c_2 \gg 1$.  Inserting the mode functions, we can get the relations for sub-horizon and super-horizon modes respectively,
 \begin{equation}\label{P_and_Omega}
 P_{\gamma}(k,\tau_{\text{obs}}\rightarrow0)=
\begin{cases}
  24 \Big( \frac{a(\tau_*)H}{k} \Big)^4   \Omega_{GW}(k,\tau_*) ,  & |k\tau_*| \gg 1
\\
 24 \Big( \frac{a(\tau_*)H}{k} \Big)^2   \Omega_{GW}(k,\tau_*) ,   & |k\tau_*| \ll 1
\end{cases} ~.
 \end{equation}

\section{Gravitational waves by the bubbles}\label{Sec_Bubble_GW_Spectrum}
The sub-horizon case in our inflationary scenarios is similar to the previous semi-numerical studies on the EW phase transitions such as \cite{Huber:2008hg,Kamionkowski:1993fg}. Usually the phase transition is a rapid process compared to the Hubble time and thus the effect of expansion of the universe can be ignored even during inflation. 

We can then finally determine the power spectrum of GW generated by bubbles. Using Eq.~(\ref{P_and_Omega}), we arrive at:
\begin{equation}\label{final_powerS}
P_{\gamma}(k,\tau_{\text{obs}}\rightarrow 0 )=P_{\gamma}^{\text{crit}}  \Big( \frac{k_{\text{crit}}}{k}\Big)^4\frac{(a+b) k_{\text{crit}} ^b k^a}{ b k_{\text{crit}} ^{a+b} +a k^{a+b} }~,
\end{equation}
where $P_{\gamma}^{\text{crit}}$ is the power spectrum at the critical point:
$
P_{\gamma}^{\text{crit}} =24   \Big(\frac{a(\tau_*)H}{k_{\text{crit}}}\Big)^4  \Omega_{GW}^{\text{crit}}~.
$
An estimation yields: $P_{\gamma}^{\text{crit}}\sim\Big(\frac{H}{\beta}\Big)^6
   \Big( \frac{\rho_{\text{higgs}} }{\rho_{\text{tot}}}\Big)^2$ for the sub-horizon case.

\begin{figure}[h]
\begin{center}
\includegraphics[width=3.5in]{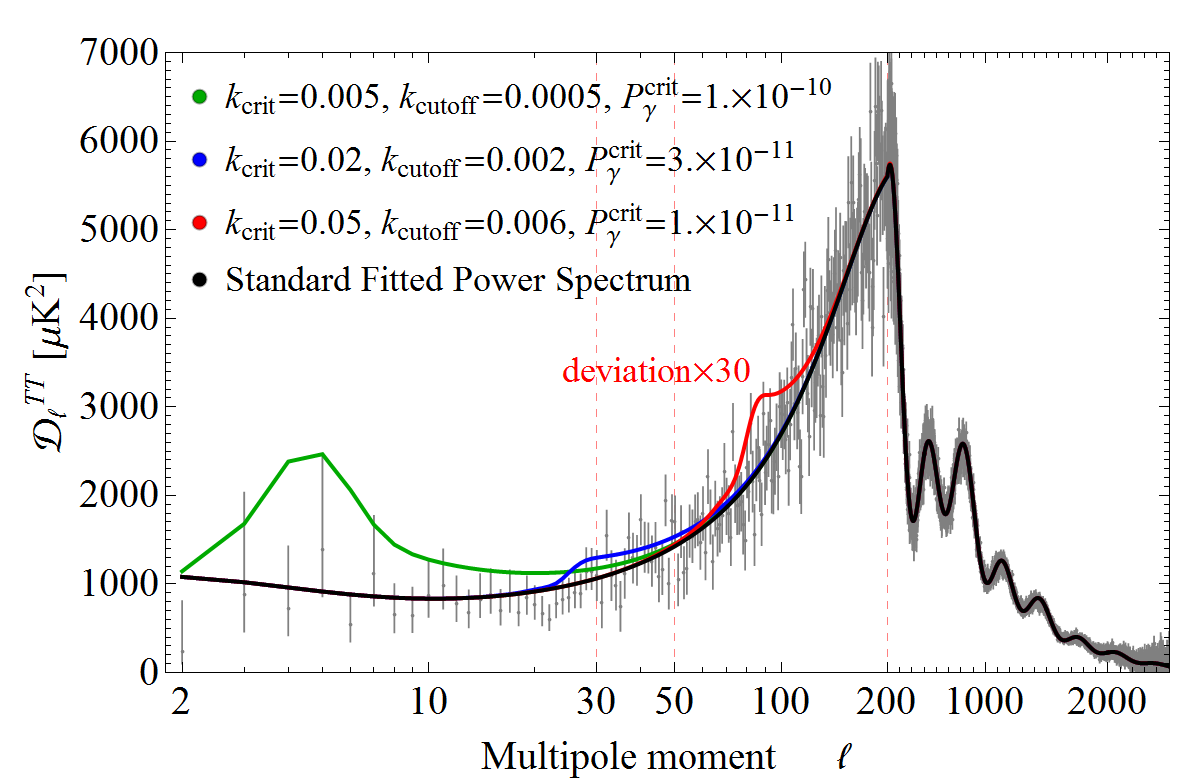}
\end{center}
\caption{CMB temperature fluctuation power spectrum. The gray  points and error bars are the experimental data of \emph{Planck} 2015 while the black curve is the fitted CMB spectrum  using the \emph{Planck} 2015 parameters.(The unit of the momentum $k$ is Mpc$^{-1}$.).}
\label{aba:fig1}
\end{figure}

\begin{figure}[h]
\begin{center}
\includegraphics[width=3.5in]{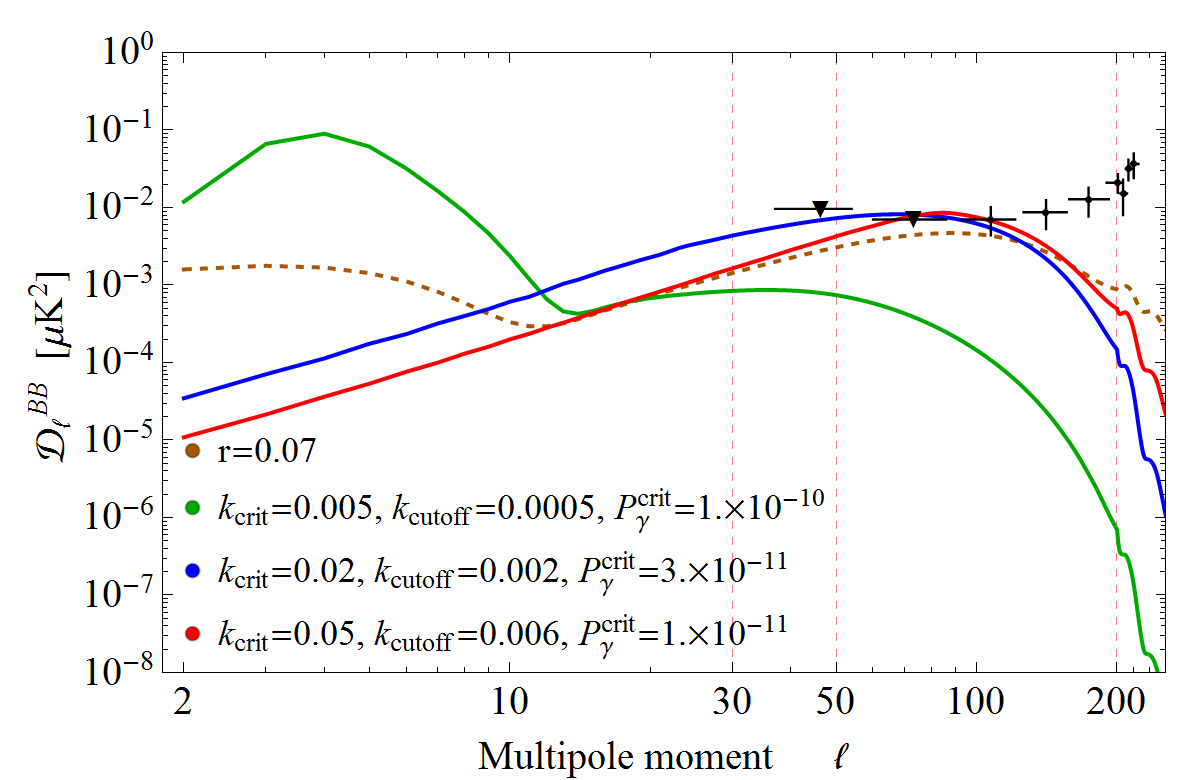}
\end{center}
\caption{B-mode power spectrum generated by phase transition bubbles.  For comparison, the primordial GWs caused by quantum vacuum fluctuations are showed in dashed line with tensor-to-scalar ratio $r=0.07$, which is the strongest upper bound to date on primordial GWs. The black ones are the CMB component bandpowers obtained from BICEP2 \& {\it Keck  Array}  experiment. }
\label{aba:fig1}
\end{figure}

 \section{Imprints on CMB}\label{Sec_CMBimprint}

The CMB spectrum can be obtained by inputting the power spectrum into the CLASS \cite{Blas} where the transfer function is calculated. As we see from Eq.~(\ref{final_powerS}), the power spectrum diverges when $k\rightarrow 0$. This divergence is unphysical as the GW generating formulas break down at super horizon scale. 
We can introduce the horizon scale as a natural cut-off, yielding $k_{\text{cutoff-physical}}=k_{\text{cutoff}}/a_* \sim H $. The critical physical momentum is related to the bubble size via $k_{\text{crit} }/a_*\sim R_b^{-1}$. The corresponding comoving momenta are
$
k_{\text{crit}} \sim \frac{1}{v_b}   \frac{\beta}{H}    e^{N_*}     k_0, \quad
k_{\text{cutoff}} \sim   e^{N_*}     k_0,
$
where $N_*$ is the e-folding number of phase transition counting from the time that the largest mode $k_0$ exits the horizon.  We choose the scale factor  today $a_0$ as one, thus the largest physical mode today is $k_0=0.0002 \text{Mpc}^{-1}$ as the inverse of observed universe size. Approximately we can find $k_0 $ corresponding to the position at CMB multipole $\ell_0\sim 2$. Then we arrive at relations: $\ell_{\text{crit}} \sim 2e^{N_*}\beta/(v_b H  )   ,\ell_{\text{cutoff}} \sim  2 e^{N_*}$.
\begin{itemize}
\item GUT scenarios: The GUT phase transition happens around $10^{16}$ GeV \cite{Georgi:1974yf,Guth:1979bh}. If choosing $(H/\beta)^6 \sim 10^{-10}, \rho_\text{GUT} \sim (10^{16}\text{GeV})^4$, and $H  \sim 10^{14}\text{GeV} $, then we have  $P_{\gamma}^{\text{crit}}\sim 10^{-10}$.  
\item EW scenarios: Typically the EWPT temperature is around $10^2 \text{GeV} \sim 10^3 \text{GeV}$, which requires $T_{GH} \sim H< 10^2 \text{GeV}$. A first-order EWPT can be achieved in various theories beyond the standard model, e.g. \cite{Cohen:1993nk}. If taking $(H/\beta)^6 \sim 10^{-6}, \rho_\text{higgs} \sim (10^3\text{GeV})^4$, and $H \sim 10^{-11} \text{GeV} $, then we have  $P_{\gamma}^{\text{crit}}\sim 10^{-10}$. The baryogenesis requires the universe going out of equilibrium, sometimes a strong first-order phase transition. Although the phase transition during inflation is not suited for generating baryons because all the particle generated during inflation will be diluted. The second symmetry restoration (EWPTb or a contrived GUTb) we proposed here might achieve that goal(see a related model \cite{Dolgov:1996qq}).
\end{itemize}

\section{Discussions}\label{Sec_conclusion}
In this talk we introduce an indirect approach to probe the first-order CPTs, such as the GUT  phase transition and the EWPT, by detecting the GWs through the CMB. These GWs are generated during inflation via bubble collisions or by the turbulence caused by bubble motion in the thermal plasma, characterized with a scale-dependent power spectrum. 

The large-scale scalar power spectrum caused by the first-order CPTs during inflation might be suppressed due to the relatively large values of the slow-roll parameter when the thermal radiation is diluted. The further discussion on the density fluctuation depends on the details of the specific inflation models.

\section*{Acknowledgments}
Sichun Sun was supported by the CRF Grants of the Government of the Hong Kong SAR under HUKST4/CRF/13G.


\begin{thebibliography}{10}

\bibitem{Jiang:2015qor} 
  H.~Jiang, T.~Liu, S.~Sun and Y.~Wang,
  ``Echoes of Inflationary Particle Phase Transitions in the CMB'',
  arXiv:1512.07538 [astro-ph.CO].
  
  
\bibitem{Hamada:2015skp} 
  Y.~Hamada, T.~Noumi, S.~Sun and G.~Shiu,
  ``An O(750) GeV Resonance and Inflation,''
  arXiv:1512.08984 [hep-ph].


  \bibitem{Bezrukov:2007ep}
  F.~L.~Bezrukov and M.~Shaposhnikov,
  ``The Standard Model Higgs boson as the inflaton,''
  Phys.\ Lett.\ B {\bf 659} (2008) 703
  ,arXiv:0710.3755 [hep-th];
  D.~Buttazzo, G.~Degrassi, P.~P.~Giardino, G.~F.~Giudice, F.~Sala, A.~Salvio and A.~Strumia,
  ``Investigating the near-criticality of the Higgs boson,''
  JHEP {\bf 1312} (2013) 089
  ,arXiv:1307.3536 [hep-ph].
  
\bibitem{Coleman:1980aw} 
  S.~R.~Coleman and F.~De Luccia,
  Phys.\ Rev.\ D {\bf 21}, 3305 (1980).
  doi:10.1103/PhysRevD.21.3305
  
  \bibitem{Blas} Blas, D., Lesgourgues, J.,
\& Tram, T., JCAP, 7, 034 (2011) [arXiv:1104.2933 [astro-ph]]
  
\bibitem{Huber:2008hg}
  S.~J.~Huber and T.~Konstandin,
  ``Gravitational Wave Production by Collisions: More Bubbles,''
  JCAP {\bf 0809}, 022 (2008)
  [arXiv:0806.1828 [hep-ph]].
  
  \bibitem{Gibbons:1977mu}
  G.~W.~Gibbons and S.~W.~Hawking,
  Phys.\ Rev.\ D {\bf 15}, 2738 (1977).
  
\bibitem{Kamionkowski:1993fg}
  M.~Kamionkowski, A.~Kosowsky and M.~S.~Turner,
  ``Gravitational radiation from first order phase transitions,''
  Phys.\ Rev.\ D {\bf 49}, 2837 (1994)
  [astro-ph/9310044].
  
\bibitem{Georgi:1974yf}
  H.~Georgi, H.~R.~Quinn and S.~Weinberg,
  Phys.\ Rev.\ Lett.\  {\bf 33}, 451 (1974).


\bibitem{Guth:1979bh}
  A.~H.~Guth and S.~H.~H.~Tye,
  ``Phase Transitions and Magnetic Monopole Production in the Very Early Universe,''
  Phys.\ Rev.\ Lett.\  {\bf 44}, 631 (1980)
  [Phys.\ Rev.\ Lett.\  {\bf 44}, 963 (1980)].
  
\bibitem{Cohen:1993nk}
  A.~G.~Cohen, D.~B.~Kaplan and A.~E.~Nelson,
  ``Progress in electroweak baryogenesis,''
  Ann.\ Rev.\ Nucl.\ Part.\ Sci.\  {\bf 43}, 27 (1993)
  [hep-ph/9302210].
  
\bibitem{Dolgov:1996qq} 
  A.~Dolgov, K.~Freese, R.~Rangarajan and M.~Srednicki,
  ``Baryogenesis during reheating in natural inflation and comments on spontaneous baryogenesis,''
  Phys.\ Rev.\ D {\bf 56}, 6155 (1997)
  doi:10.1103/PhysRevD.56.6155
  [hep-ph/9610405].



\end{thebibliography}
\end{document}